\documentclass[11pt]{article}
\usepackage[utf8]{inputenc}
\usepackage[T1]{fontenc}
\usepackage{lmodern}
\usepackage[margin=1in]{geometry}
\usepackage{amsmath,amssymb,amsthm}
\usepackage{graphicx}
\usepackage{booktabs}
\usepackage{array}
\usepackage{xcolor}
\usepackage{microtype}
\usepackage{caption}
\usepackage{subcaption}
\usepackage{float}
\usepackage{flafter}
\usepackage{placeins}
\usepackage[hidelinks]{hyperref}
\usepackage{url}
\newcommand{\snr}{\mathrm{snr}}
\newcommand{\cin}{c_{\mathrm{in}}}
\newcommand{\cout}{c_{\mathrm{out}}}
\newcommand{\VI}{\mathrm{VI}}
\newcommand{\E}{\mathbb{E}}
\newcommand{\1}{\mathbf{1}}

\title{\vspace{-1.5em}Detector-Output Instability near the Kesten--Stigum Boundary:\\
Separating Hard Readout, Relaxation, and Fixed-Point Dispersion}
\author{Faruk Alpay\thanks{Corresponding author: \texttt{alpay@lightcap.ai}} \quad Barış Başaran\\[3pt]
  \small Department of Computer Engineering, Bahçeşehir University, Istanbul, Turkey\\[-1pt]
  \small \texttt{\{faruk.alpay, baris.basaran\}@bahcesehir.edu.tr}}
\date{}

\begin{document}
\maketitle

\begin{abstract}
Community-detection algorithms usually return a single partition, even when the
same network supports several plausible outputs under independent initializations
or small data perturbations. We study this output distribution through three
paired observables: hard-partition variation of information (VI), a residual-gated
fixed-point VI, and a cutoff-free Jensen--Shannon distance between BP marginal
fields. For the symmetric sparse stochastic block model, linearizing belief
propagation around the uninformative fixed point gives the standard
Kesten--Stigum onset condition $(\cin-\cout)^2/(q^2c)=\snr^2=1$, with
$\snr=(\cin-\cout)/(q\sqrt c)$. This calculation locates the linear onset. The
maximum of the hard VI readout is a finite-size detector curve: in the synthetic
sweeps it lies on the detectable side, typically $\snr^\star\simeq1.05$--$1.10$.
The hard-readout peak is readout-dependent. On the same BP outputs, changing the
polarization cutoff from $0.001$ to $0.1$ moves $\snr^\star$ across
$1.047$--$1.128$ and lowers the peak VI from about $1.83$ to $1.49$ bits at the
largest cutoff. The nontrivial-readout activation point has a clean cutoff law:
the median activation offset satisfies
$\snr_{50}(\tau)-1=0.0086+0.522\tau$ with $R^2=0.996$ over the same sweep.
Long-budget residual gating separates this readout effect from critical slowing.
At $\snr=1.05$ and $1.10$, the 2500-iteration hard VI is
$1.49$ and $1.58$ bits, but the residual-gated subsets have zero VI and zero
marginal dispersion. From $\snr=1.15$ through $1.30$, all or nearly all runs pass
the gate and retain VI $1.31$ down to $1.24$ bits, with marginal dispersion
$0.056$--$0.085$ bits per node. A high-replication large-$N$ audit uses thirty
graph draws at each sampled point through $N=100000$. The large-$N$ peak
locations are $1.024$, $1.022$, $1.025$, and $1.024$ for
$N=16000,32000,64000,100000$. A zero-asymptote power law is disfavored; the
measured high-$N$ behavior is a small plateau
$\snr^\star-1\simeq0.024$ with graph-bootstrap 90\% interval
$[0.0227,0.0316]$. On real networks, a label-free Bethe-Hessian modularity margin
and a Chung--Lu null gate are evaluated on political blogs and six SNAP graphs.
The real-network benchmark confirms that the measurement can be run label-free,
while heterogeneous networks can retain null-significant structure even after
strong edge subsampling. The final claim is a
detector-output decomposition near the Kesten--Stigum boundary, with hard
readout, relaxation dynamics, and fixed-point-field dispersion reported
separately.
\end{abstract}

\section{Introduction}

Community detection asks for a partition of a network into groups that are densely
connected internally and sparsely connected externally~\cite{fortunato2010,
fortunato2016}. Standard algorithms return one partition, but the inferential
object is often a distribution. Independent optimizer initializations, bootstrap
edge perturbations, and changes of objective can yield partitions that have
similar scores but different node assignments. The issue goes beyond
implementation: modularity maximization has an extensively degenerate landscape of
high-scoring partitions~\cite{good2010}, and recent work on solution-landscape
exploration and overlap-gap geometry treats the distribution of near-optimal
partitions as a structural feature of the problem~\cite{calatayud2019,
bhamidi2026}. Related ensemble perspectives appear in consensus clustering and
Bayesian blockmodel uncertainty~\cite{lancichinetti2012,peixoto2019}. We call the
dispersion of plausible algorithmic outputs \emph{partition degeneracy}.

The stochastic block model (SBM) supplies a controlled setting in which this
dispersion can be related to a known transition. In the sparse symmetric SBM,
community recovery undergoes the Kesten--Stigum transition~\cite{kesten1966,
decelle2011prl,decelle2011pre,massoulie,mossel2015,abbe2017}: above the threshold
the planted partition has nonzero overlap with efficient estimators, whereas
below it no estimator improves over chance. Belief propagation attains this
threshold~\cite{decelle2011pre}; spectral algorithms based on the non-backtracking
operator and the Bethe Hessian attain the same boundary~\cite{krzakala2013,
saade2014}. Finite networks round the transition but preserve its scaling
structure~\cite{young2017}. The statistical-physics description is a changing
landscape of fixed points and metastable states. Kawamoto and Kabashima counted
metastable modularity states and located an algorithmic limit through the
complexity of that landscape~\cite{kawamoto2019}.

This paper turns that landscape picture into a detector-output decomposition. We
measure the hard partition readout used by a practical pipeline, the
cutoff-free dispersion of BP marginal fields, and the residual-gated part of the
BP ensemble that has reached a fixed point. The hard VI curve has an offset peak
above $\snr=1$ in finite sparse SBMs. The new controls show how much of that peak
comes from hard readout and critical slowing, and where a budget-stable
fixed-point ensemble remains.

\paragraph{Contributions and scope.} The paper makes the following claims and
limitations explicit.
\begin{itemize}\itemsep2pt
\item We define hard-readout partition degeneracy as the mean variation of
information among an algorithmic ensemble of detected partitions and express it
as a function of the partition-overlap matrix. The definition is
label-permutation invariant and requires no ground truth, but it is a joint
property of the detector ensemble and the graph
(Sections~\ref{sec:obs}--\ref{sec:math}).
\item We recall the Kesten--Stigum scaling coordinate from the linearization of
belief propagation around the uninformative fixed point. This calculation gives
the onset reference $\snr=1$. We use cutoff-free marginal-field dispersion and
residual-gated VI to separate hard readout from BP fixed-point convergence
(Section~\ref{sec:math}).
\item We measure the finite-size offset of the hard-readout ridge. In the phase
diagram it lies in $1.05\le\snr^\star\le1.15$, and in the size sweep it moves
from $\snr^\star\approx1.09$ at $N=500$ to $\snr^\star\approx1.05$ at $N=8000$.
We add a high-replication large-$N$ audit through $N=100000$. The large-$N$
offset is a small plateau near $0.024$, so the earlier zero-asymptote exponent
fit is replaced by a plateau scaling statement
(Sections~\ref{sec:phase} and~\ref{sec:fss}).
\item We report a peak-normalized alignment across twelve block-count and
mean-degree combinations and compare its spread to a permutation null. This
supports a nonrandom alignment of the finite-size detector curves while staying
below an exact universality claim (Section~\ref{sec:collapse}).
\item We use a partition-space embedding to show that the detected-partition cloud
broadens in the weakly detectable band, and we add controls that separate
fixed-graph initialization noise from pooled graph sampling, expose the
polarization-cutoff readout artifact, record BP iteration counts, residuals,
matched-budget partition drift, a cutoff-free marginal-dispersion observable, and
a Bethe free-entropy proxy (Sections~\ref{sec:landscape}--\ref{sec:controls}).
\item We add a $q=5$ hard-phase probe comparing random BP initialization with a
weak planted warm start. The probe finds no well-converged sub-Kesten--Stigum
coexistence band for the tested detector, so the main claims remain restricted to
the continuous $q\le4$ regime (Section~\ref{sec:hard}).
\item We repeat the measurement on political blogs and on six public SNAP
networks. The real-network coordinate is label-free: a Bethe-Hessian modularity
margin with a Chung--Lu null gate. Political labels enter only as an external
recovery check for the blogs graph (Section~\ref{sec:real}).
\end{itemize}

\section{The degeneracy observable}
\label{sec:obs}

\paragraph{Model and threshold.} We work with the symmetric stochastic block
model~\cite{holland1983}: $n$ nodes are split into $q$ equal groups, two nodes in
the same group are joined with probability $\cin/n$ and two nodes in different
groups with probability $\cout/n$, so the mean degree is
$c=(\cin+(q-1)\cout)/q$. The detectability transition sits at the
Kesten--Stigum threshold $\cin-\cout=q\sqrt{c}$~\cite{decelle2011pre,mossel2015}.
We use the dimensionless detectability driver
\begin{equation}
\snr=\frac{\cin-\cout}{q\sqrt{c}},
\end{equation}
so the threshold is at $\snr=1$ for every $q$ and $c$, with $\snr<1$ undetectable
and $\snr>1$ detectable. For $q\le 4$ the algorithmic and information-theoretic
thresholds coincide and the transition is continuous~\cite{decelle2011pre}; we
restrict the synthetic study to that regime.

\paragraph{Detector.} On synthetic graphs, where the parameters are known, we
detect with belief propagation, the model-aware message-passing algorithm that
realizes the Kesten--Stigum threshold~\cite{decelle2011pre}. Each directed edge
$i\to j$ carries a message $\psi^{i\to j}_t$, the probability that $i$ is in group
$t$ in the cavity where $j$ is removed, updated by
\begin{equation}
\psi^{i\to j}_{t}\;\propto\;\prod_{k\in\partial i\setminus j}
\sum_{s} c_{ts}\,\psi^{k\to i}_{s},
\label{eq:bp}
\end{equation}
with $c_{ts}=\cin$ for $t=s$ and $\cout$ otherwise. We start from random messages
and project out the net magnetization each sweep, the instantaneous form of the
Decelle external field, which keeps the iteration off the trivial ferromagnetic
mode. When belief propagation converges to the uninformative fixed point, the node
marginals are uniform. The hard partition readout uses a polarization cutoff
($0.02$ in the main synthetic sweeps) and returns the trivial single-block
partition for low-polarization fields. This rule removes arbitrary hard labels
from uniform marginals, while it also fixes the left baseline of the hard VI
curve. The measured hard-readout peak combines three effects: the onset of
informative BP fixed points, the cutoff that maps weakly polarized outputs to the
trivial partition, and the concentration of detected partitions deeper in the
detectable phase.

\paragraph{The observable.} For a fixed set of model parameters we form an
ensemble of detected partitions by combining independent graph draws with
independent BP initializations. On the real network, where only one graph is
available at each dilution level, the data perturbation is an edge-subsampling
bootstrap. The partition degeneracy is the mean variation of
information~\cite{meila2007} among the ensemble,
\begin{equation}
D=\E_{P,P'\sim\mu_G}\big[\VI(P,P')\big],
\qquad
\VI(P,P')=H(P)+H(P')-2I(P,P'),
\label{eq:degeneracy}
\end{equation}
where $H$ is the partition entropy and $I$ the mutual information, measured in
bits, and $\mu_G$ is the empirical distribution induced by the detector,
initialization protocol, and graph perturbation scheme. Variation of information
is a metric and is invariant to relabeling, so a partition and its group-permuted
copy have distance zero. Unless a control explicitly separates the components, we
pool the partitions before averaging; this measures the dispersion of outputs a
practitioner would see from both algorithmic and sampling variability. The
observable is an operational detector-output distribution for a specified
detector, finite iteration budget, initialization
scheme, and perturbation scheme.

To separate the hard readout from the BP state itself, the controls also use a
cutoff-free marginal dispersion. If $\Psi$ and $\Phi$ are two BP marginal fields,
define
\begin{equation}
D_\Psi(\Psi,\Phi)
  = \min_{\pi\in S_q}\frac1n\sum_{i=1}^n
    \mathrm{JS}\!\left(\Psi_i,\pi\Phi_i\right),
\label{eq:marginal-js}
\end{equation}
where $\mathrm{JS}$ is the Jensen--Shannon divergence in bits and the minimum
removes arbitrary group-label permutations. The ensemble statistic is the mean
of $D_\Psi$ over pairs of runs. Uniform BP marginals give $D_\Psi=0$ with zero
polarization-cutoff dependence, so this control tests whether the BP fixed-point
fields themselves differ, beyond the hard threshold that maps them to the trivial
partition.

Uniform marginal fields have zero $D_\Psi$. If one samples hard labels
independently from those same uniform marginals, the empirical VI tends to
$2\log_2 q$; this elementary fact is used only as a readout sanity check in
Section~\ref{sec:controls}. For fixed-point claims we also use a residual-gated
statistic. Let $r(\Psi)$ be the undamped BP residual of the converged message
field and set
\begin{equation}
D_{\epsilon}^{\mathrm{fp}}
  = \E\!\left[\VI(P,P')\mid r(\Psi)\le\epsilon,\ r(\Phi)\le\epsilon\right],
\qquad
D_{\Psi,\epsilon}^{\mathrm{fp}}
  = \E\!\left[D_\Psi(\Psi,\Phi)\mid r(\Psi)\le\epsilon,\ r(\Phi)\le\epsilon\right].
\label{eq:residual-gated}
\end{equation}
Runs outside the gate are relaxation data. We report the gated quantity only
when the gate contains enough runs for pairwise distances.

\section{Threshold reference and partition-overlap geometry}
\label{sec:math}

The next calculations fix the reference point for the measurement. They are
included to make the scaling coordinate and the VI observable explicit; by
themselves they leave the finite-size maximum, offset, and amplitude of the
hard-readout curve. The first calculation identifies the linear onset. Write the BP message
on a directed edge as a small perturbation of the uninformative fixed point,
\begin{equation}
\psi^{i\to j}_t=\frac1q+\eta^{i\to j}_t,\qquad
\sum_{t=1}^q\eta^{i\to j}_t=0 .
\end{equation}
For the symmetric SBM the affinity matrix can be decomposed as
\begin{equation}
C=\cout \mathbf{1}\mathbf{1}^{\mathsf T}
  +(\cin-\cout)I_q .
\end{equation}
Linearizing Eq.~\eqref{eq:bp} around the uninformative fixed point and projecting
onto the $(q-1)$-dimensional subspace orthogonal to $\mathbf 1$ gives
\begin{equation}
\eta^{i\to j}_{t,+}
  = \frac{\cin-\cout}{qc}
    \sum_{k\in\partial i\setminus j}\eta^{k\to i}_{t}
    +O(\|\eta\|^2).
\label{eq:linear-bp}
\end{equation}
Thus the linearized dynamics is the non-backtracking operator multiplied by
$\lambda=(\cin-\cout)/(qc)$. On a locally tree-like sparse SBM, a perturbation
has roughly $c^\ell$ non-backtracking descendants at depth $\ell$, while each
message correlation is multiplied by $\lambda^\ell$. The second moment therefore
evolves as
\begin{equation}
\E\|\eta_\ell\|^2 \asymp (c\lambda^2)^\ell\E\|\eta_0\|^2
 = \snr^{2\ell}\E\|\eta_0\|^2,
\qquad
\snr=\frac{\cin-\cout}{q\sqrt c}.
\label{eq:ks-growth}
\end{equation}
The uninformative fixed point is linearly stable for $\snr<1$ and unstable for
$\snr>1$, which is the Kesten--Stigum condition. This derivation gives the onset
reference used throughout the paper. The observed VI peak is a nonlinear,
finite-size property of the detector ensemble and can be displaced from this
onset.

The second calculation makes precise what the VI ensemble measures. For two
partitions $P$ and $P'$, define their empirical overlap matrix
\begin{equation}
Q_{ab}(P,P')=\frac1n\sum_{i=1}^n \1\{P_i=a,\;P'_i=b\},\qquad
\pi_a=\sum_b Q_{ab},\qquad \pi'_b=\sum_a Q_{ab}.
\end{equation}
Then
\begin{equation}
\VI(P,P')
=-\sum_a\pi_a\log_2\pi_a-\sum_b\pi'_b\log_2\pi'_b
-2\sum_{ab}Q_{ab}\log_2\frac{Q_{ab}}{\pi_a\pi'_b}.
\label{eq:vi-q}
\end{equation}
Equation~\eqref{eq:vi-q} shows that $D$ is the first moment of the
partition-overlap distribution generated by $\mu_G$. It is zero in two distinct
limits: when $\mu_G$ concentrates on the trivial one-block partition, and when it
concentrates on one informative partition up to label permutation. It is large
only when $\mu_G$ has substantial mass on mutually distant partitions. The
hard-readout peak is therefore a width measure for the algorithmic output
distribution, separate from monotone recoverability; it is a
measure of the width of the algorithmic output distribution in the weakly
detectable band.

Below threshold the linear susceptibility of the BP fixed point has the geometric
form
\begin{equation}
\chi_{\mathrm{lin}}(\snr)
  \propto \sum_{\ell\ge0}\snr^{2\ell}
  = \frac{1}{1-\snr^2},\qquad \snr<1,
\label{eq:linear-susceptibility}
\end{equation}
before finite-size and nonlinear terms round the divergence. The empirical
susceptibility in Section~\ref{sec:fss} uses the planted labels only to verify
this critical scaling; Eq.~\eqref{eq:degeneracy} is label-free.

The same linearization also explains why convergence diagnostics are necessary.
Loopy BP fixed points and their Bethe free-energy interpretation are standard
objects in the message-passing literature~\cite{yedidia2003,heskes2003,mooij2007};
near a loss of stability, residuals and relaxation times must be recorded before
interpreting a hard partition ensemble as a fixed-point ensemble.
With damping parameter $\alpha$, the leading informative perturbation of the
implemented update contracts approximately as
\begin{equation}
a_{t+1}\simeq \rho_\alpha(\snr)a_t,\qquad
\rho_\alpha(\snr)=\alpha+(1-\alpha)\snr ,
\label{eq:damped-rate}
\end{equation}
on the stable side of the transition. Hence the linear relaxation time scales as
\begin{equation}
\tau_{\mathrm{rel}}
  \simeq \frac{1}{1-\rho_\alpha(\snr)}
  = \frac{1}{(1-\alpha)(1-\snr)} ,
\qquad \snr<1,
\label{eq:relaxation-time}
\end{equation}
before finite-size rounding cuts off the divergence. A finite iteration budget
therefore has its largest effect precisely near the Kesten--Stigum point. By
itself this mechanism leaves subthreshold outputs trivial, while it can broaden
the detector-output distribution on the weakly detectable side.

A minimal finite-size model shows why the VI maximum can move away from the
linear onset. Let $T$ be the trivial one-block partition and suppose the detector
readout near threshold is a mixture
\begin{equation}
\mu_{\snr}=\tau(\snr)\delta_T+(1-\tau(\snr))\nu_{\snr},
\end{equation}
where $\nu_{\snr}$ is the distribution of nontrivial BP outputs that pass the
polarization cutoff. Since $\VI(T,P)=H(P)$ for every nontrivial partition $P$,
the hard-readout VI decomposes as
\begin{equation}
D(\snr)
 =2\tau(1-\tau)\,\E_{\nu_{\snr}}H(P)
 +(1-\tau)^2\E_{\nu_{\snr},\nu_{\snr}}\VI(P,P').
\label{eq:mixture-degeneracy}
\end{equation}
The first term is largest when the detector alternates between the trivial
readout and weakly polarized informative readouts; the second term measures
dispersion among the informative readouts themselves. Because the cutoff requires
finite polarization, the transition of $\tau(\snr)$ occurs on the detectable
side at finite $n$. Equation~\eqref{eq:mixture-degeneracy} gives a mechanism for
$\snr^\star$, with the peak as the maximum width of
this detector-induced mixture, distinct from the Kesten--Stigum onset
itself.

\section{Simulation protocol and validation}
\label{sec:engine}

All synthetic measurements use the BP detector described above with a fixed
iteration budget, damping, random initializations, and the reported polarization
cutoff. A parallel implementation is used only to run many independent detector
instances; the implementation backend sits outside the scientific claim. A NumPy
reference implements the same update equations, and the accelerated code is
checked against it at the level used in the paper: the final discrete partition.
On identical validation graphs the two implementations return the same partition
above threshold and both return the trivial partition below it; intermediate
floating-point messages may differ at bit level.

\begin{figure}[H]
\centering
\includegraphics[width=\linewidth]{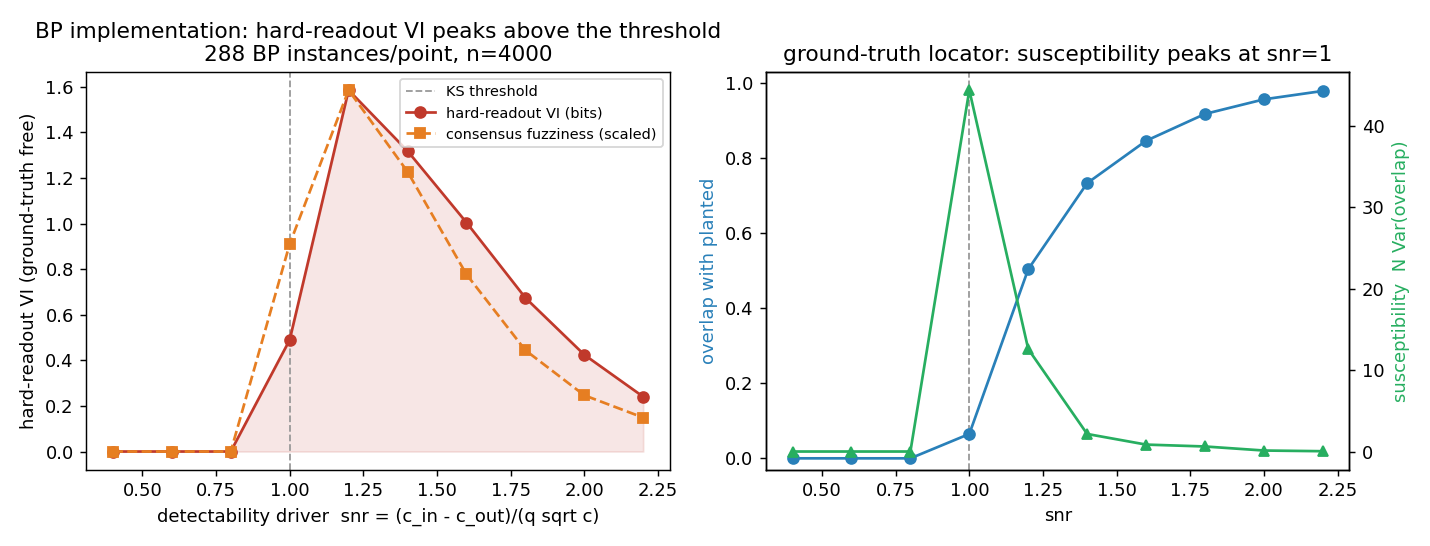}
\caption{The accelerated belief-propagation implementation reproduces the
finite-size hard-readout offset. Left: the ground-truth-free hard VI readout
(variation of information among detected partitions, with the relabeling-invariant
consensus fuzziness rescaled for comparison) peaks above the Kesten--Stigum
threshold. Right: the recovery overlap rises through the threshold and the
ground-truth susceptibility $N\,\mathrm{Var}(\text{overlap})$ peaks at $\snr=1$,
locating the transition. The accelerated implementation and the NumPy reference
agree on the final discrete partitions in the validation cases.}
\label{fig:validate}
\end{figure}

The validation sweep in Figure~\ref{fig:validate} uses $n=4000$, $q=2$, $c=10$,
six independent graph draws, and forty-eight BP restarts per graph at each of ten
$\snr$ values. The phase-ridge sweep uses $n=2500$, $q=2$, sixteen mean degrees,
twenty-one $\snr$ values, six graph draws, and forty restarts per graph. The
cross-parameter sweep uses $n=3500$, twelve $(q,c)$ combinations with
$q\in\{2,3,4\}$, eighteen $\snr$ values, six graph draws, and forty restarts per
graph. The finite-size sweep uses $N=500$--$8000$ at $q=2$, $c=10$, thirty-three
$\snr$ values, eight graph draws, and twenty-four restarts per graph. The
landscape embedding uses $n=2200$, $q=3$, $c=12$ at
$\snr=0.75,1.10,2.20$ with twenty-four graph draws and twelve restarts per graph.
The fixed-graph and cutoff controls use $n=2500$, $q=2$, $c=10$ on eleven
$\snr$ values. The convergence control uses $n=1200$, $q=2$, $c=10$ on eight
$\snr$ values, three graph draws, and eight matched BP restarts per graph at
iteration budgets $100$, $300$, and $700$, and records both hard-partition VI and
the marginal Jensen--Shannon dispersion of Eq.~\eqref{eq:marginal-js}. The
readout-artifact control uses the same $n,q,c$, a $0.05$ grid over
$0.85\le\snr\le1.25$, a $700$-iteration budget, and records cutoff VI, raw argmax
VI, sampled-marginal VI, and marginal-field dispersion on the same BP runs. The
residual-gated control uses the same $n,q,c$, matched initialization seeds at
$700$ and $2500$ iterations, and gate
$r(\Psi)\le5\times10^{-5}$ in Eq.~\eqref{eq:residual-gated}. The
cutoff sweep uses the same raw BP outputs at each $\snr$ and rereads them with
polarization cutoffs $0.001,0.002,0.005,0.01,0.02,0.05,0.1$. The
fine-grid finite-size check uses $N\in\{1000,2000,4000,8000\}$, a
$0.01$-spaced grid on $1.00\le\snr\le1.16$, ten graph draws, and twenty-four BP
restarts per graph with the longer $700$-iteration BP budget. The large-$N$ audit
uses $N\in\{16000,32000,64000,100000\}$ on the grid
$1.00,1.01,\ldots,1.06,1.08,1.10$, thirty graph draws, and twelve BP restarts per
graph; graph-level replicates are saved for bootstrap peak intervals. The collapse null
uses $5000$ independent within-curve
permutations of the sampled $\snr$ coordinates. The $q=5$ hard-phase probe uses
$n=1400$, $c=20$, seven $\snr$ values, three graph draws, and five random or
weak planted-warm BP restarts per graph. The political-blogs illustration uses
five degree-preserving rewiring realizations and thirty-two edge-subsampling
detections per realization; its horizontal coordinate is the label-free
Bethe-Hessian modularity margin of the diluted graph. The multi-network real-data
benchmark uses six public SNAP graphs--email-Eu-core, Wiki-Vote, CA-GrQc,
p2p-Gnutella08, Facebook combined, and Epinions--with edge-subsampling levels
$p_{\rm keep}\in\{1,0.85,0.70,0.55,0.40,0.25,0.15,0.10,0.05\}$. Each level uses
degree-corrected Bethe-Hessian detections gated by a Chung--Lu null modularity
threshold.

\FloatBarrier

\section{The hard-readout ridge sits near the threshold}
\label{sec:phase}

\begin{figure}[H]
\centering
\includegraphics[width=\linewidth]{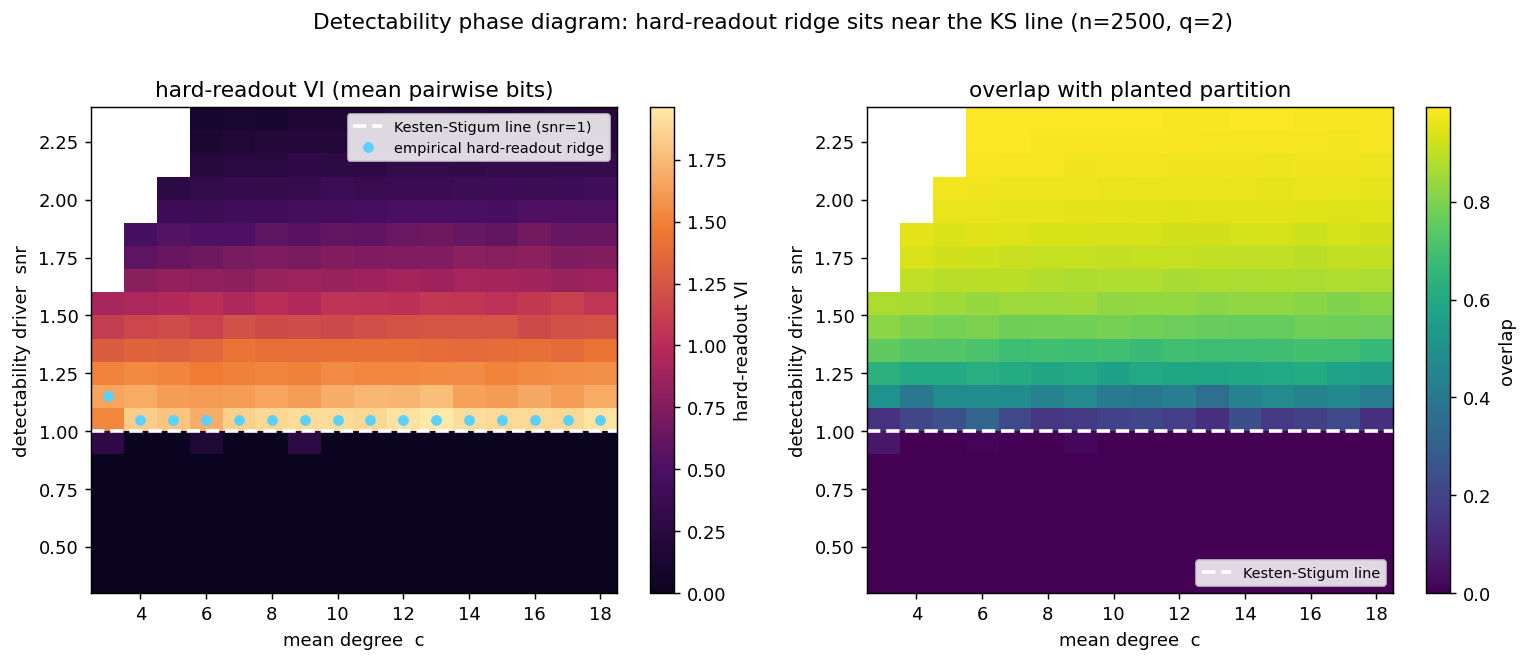}
\caption{Detectability phase diagram in the (mean degree, driver) plane. Left: the
hard-readout VI forms a ridge near the Kesten--Stigum line
($\snr=1$, dashed) at every degree; the cyan markers are the per-degree empirical
ridge. Right: the recovery overlap shows the transition at the same
line. The hard-readout VI is maximal in the thin band inside the detectable
phase.}
\label{fig:phase}
\end{figure}

We sweep the affinity plane in the coordinates of mean degree $c$ and driver
$\snr$, and at each point we average the degeneracy over several graph replicates.
The result is the phase diagram of Figure~\ref{fig:phase}. The hard-readout VI is a
ridge on the detectable side of the Kesten--Stigum line. Across the $16$ degree
values in the sweep, the empirical ridge lies in the interval
$1.05\le\snr^\star\le1.15$, with mean $1.06$. The heat map is dark on both sides:
below the line the detector returns the trivial partition and the runs agree,
whereas deep in the detectable phase the structure is strong enough that the runs
again concentrate. The companion overlap panel shows the corresponding rise of
the order parameter. The useful finite-size statement is that the detector is most unstable in a
measurably offset band where the informative fixed point is present but weakly
stable.

\FloatBarrier

\section{Coarse cross-parameter alignment}
\label{sec:collapse}

\begin{figure}[H]
\centering
\includegraphics[width=\linewidth]{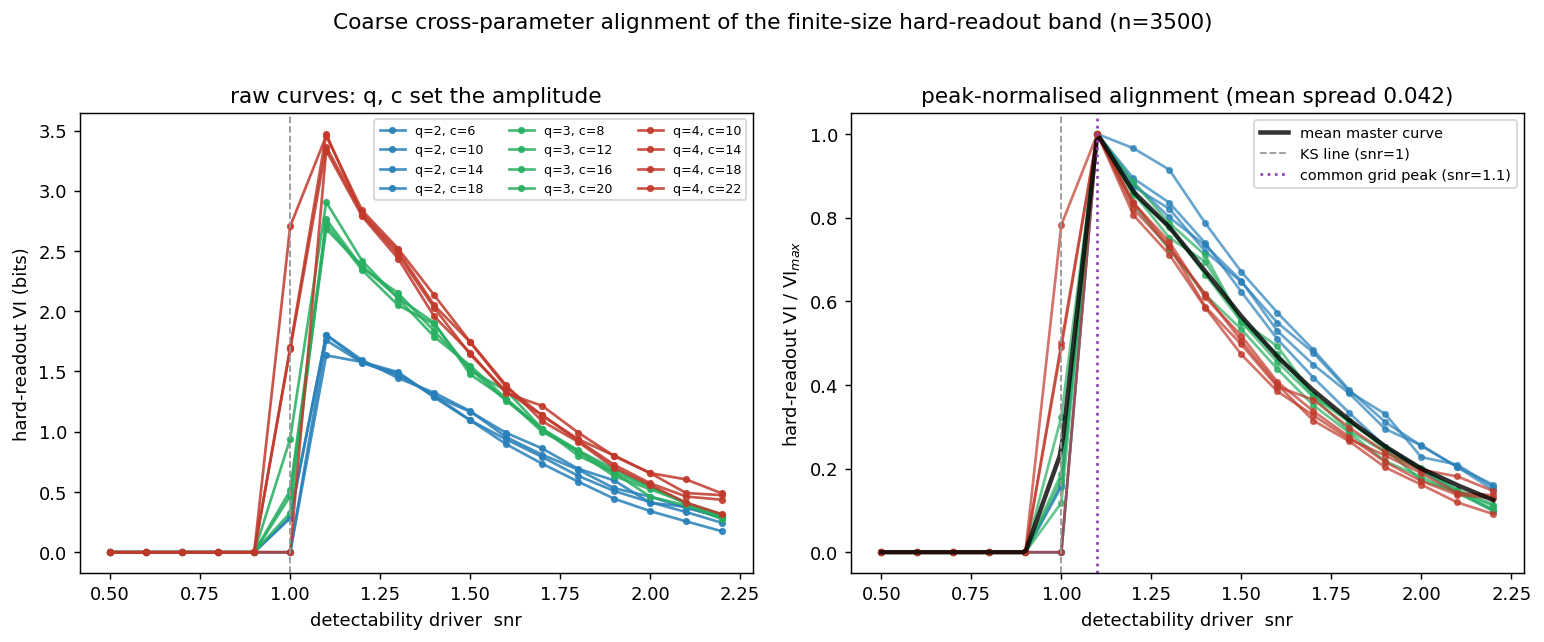}
\caption{Coarse cross-parameter alignment. Left: raw hard-readout VI
curves for twelve block-count and mean-degree combinations; all peak at
$\snr=1.10$ on this grid, with an amplitude that grows with the block count.
Right: dividing by the peak aligns the twelve curves (mean spread $0.042$). The
permutation null has mean spread $0.302$; the claim is nonrandom finite-grid
alignment below an exact universal scaling function.}
\label{fig:collapse}
\end{figure}

We next ask whether the shape of the finite-size band is similar across block
counts $q\in\{2,3,4\}$ and multiple mean degrees $c$. The expanded sweep contains
twelve curves, four for each $q$, sampled on a coarse $\snr$ grid. Every curve
has its sampled maximum at $\snr=1.10$ (Figure~\ref{fig:collapse}, left), but
this gives a coarse common peak: the grid spacing is about $0.10$,
and the finer size and cutoff sweeps already show peak motion within this range.
The peak amplitude ranges from $1.63$ bits in the $q=2$ curves to
$3.47$ bits in the $q=4$ curves. This increase is partly mechanical because VI
has a larger ceiling when there are more groups. We therefore use the raw
amplitudes only descriptively. When each curve is divided by its own peak, the
twelve curves align with a mean normalized spread of $0.042$
(Figure~\ref{fig:collapse}, right). To test whether this is only a plotting
artifact, we permuted the sampled $\snr$ coordinates independently within each
normalized curve $5000$ times. The null spread is $0.302\pm0.005$ (mean and
standard deviation), with 5--95\% interval $[0.292,0.310]$; none of the null
samples is as small as the observed spread, giving the finite-sample lower-tail
$p=2.0\times10^{-4}$. The alignment therefore reflects a shared detector-curve
shape over these coarse coordinates, below the level of an exact universal
scaling function.

\FloatBarrier

\section{Finite-size behavior of the offset}
\label{sec:fss}

\begin{figure}[H]
\centering
\includegraphics[width=\linewidth]{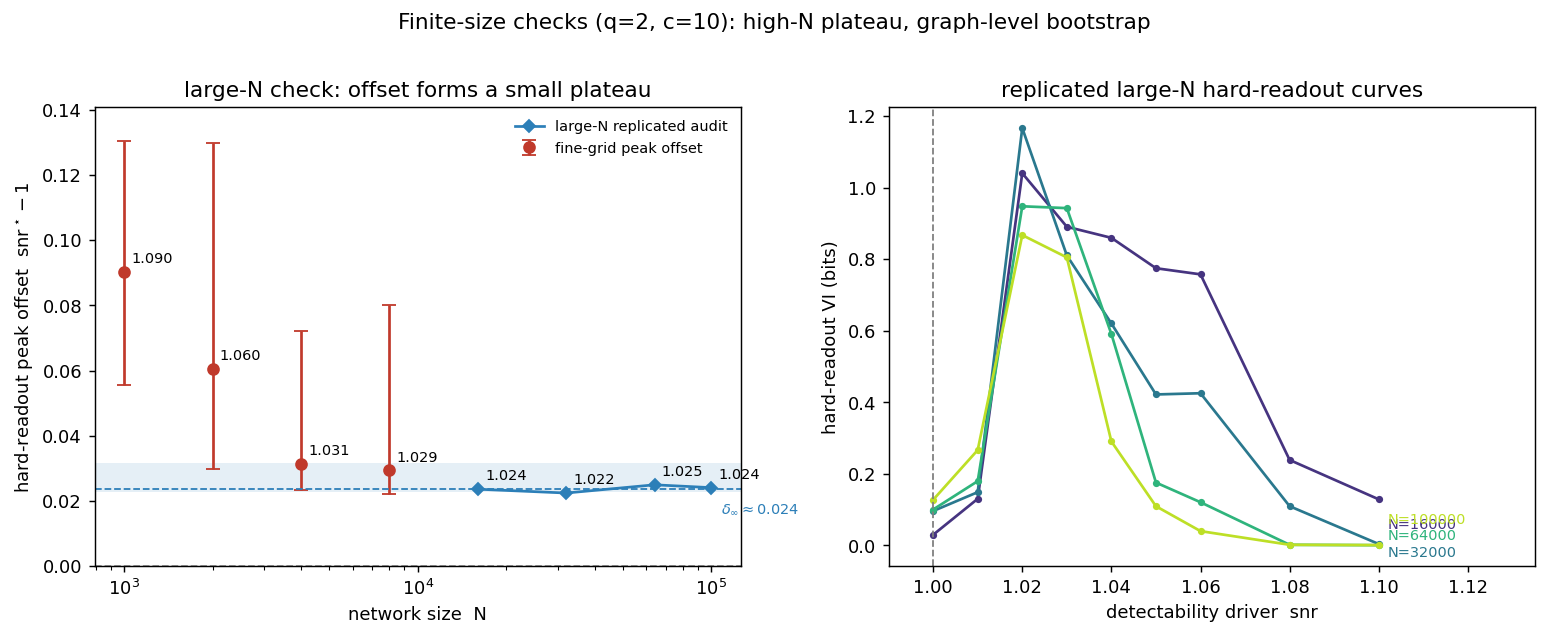}
\caption{Finite-size checks at $q=2$, $c=10$. Left: fine-grid hard-readout peak
offsets with graph-bootstrap 90\% intervals, plus the replicated large-$N$ audit
through $N=100000$ and the fitted high-$N$ plateau. Right: the replicated
large-$N$ hard-readout curves on the sampled $\snr$ grid.}
\label{fig:fss}
\end{figure}

The offset between the susceptibility maximum and the hard-readout VI maximum is
the main finite-size issue. We fix $q=2$, $c=10$ and study sizes from
$N=500$ to $N=100000$ (Figure~\ref{fig:fss}). The earlier broad sweep from
$N=500$ to $N=8000$ shows the hard-readout peak moving from
$\snr^\star\approx1.09$ at $N=500$ toward $\snr^\star\approx1.05$ at $N=8000$.
To resolve the offset near the onset, we repeated the measurement on a
$0.01$ grid over $1.00\le\snr\le1.16$ with ten graph draws per size. The
interpolated peaks are $1.090$ for $N=1000$, $1.060$ for $N=2000$, $1.031$ for
$N=4000$, and $1.029$ for $N=8000$; graph-bootstrap 90\% intervals are
respectively $[1.055,1.130]$, $[1.030,1.130]$, $[1.023,1.072]$, and
$[1.022,1.080]$.

We then repeated the large-$N$ audit with thirty graph draws at every sampled
point, a tenfold increase over the first audit. For
$N=16000,32000,64000,100000$, the measured peak locations are respectively
$1.024$, $1.022$, $1.025$, and $1.024$. The earlier $N=64000$ excursion to
$1.044$ disappears under graph-level replication. A zero-asymptote fit
$\snr^\star(N)-1=aN^{-\omega}$ gives $\omega=-0.028$ and weak explanatory power
($R^2=0.26$). A constant high-$N$ offset gives
$\delta_\infty=0.0237$ with graph-bootstrap 90\% interval
$[0.0227,0.0316]$, and a small-sample AICc diagnostic favors this plateau over
the zero-asymptote power fit. Thus the measured scaling law for this hard-readout
pipeline is a finite detector band of width about $0.024$ in $\snr$, distinct
from a resolved decay to the Kesten--Stigum point over the tested range.

\FloatBarrier

\section{The solution landscape}
\label{sec:landscape}

\begin{figure}[H]
\centering
\includegraphics[width=0.94\linewidth]{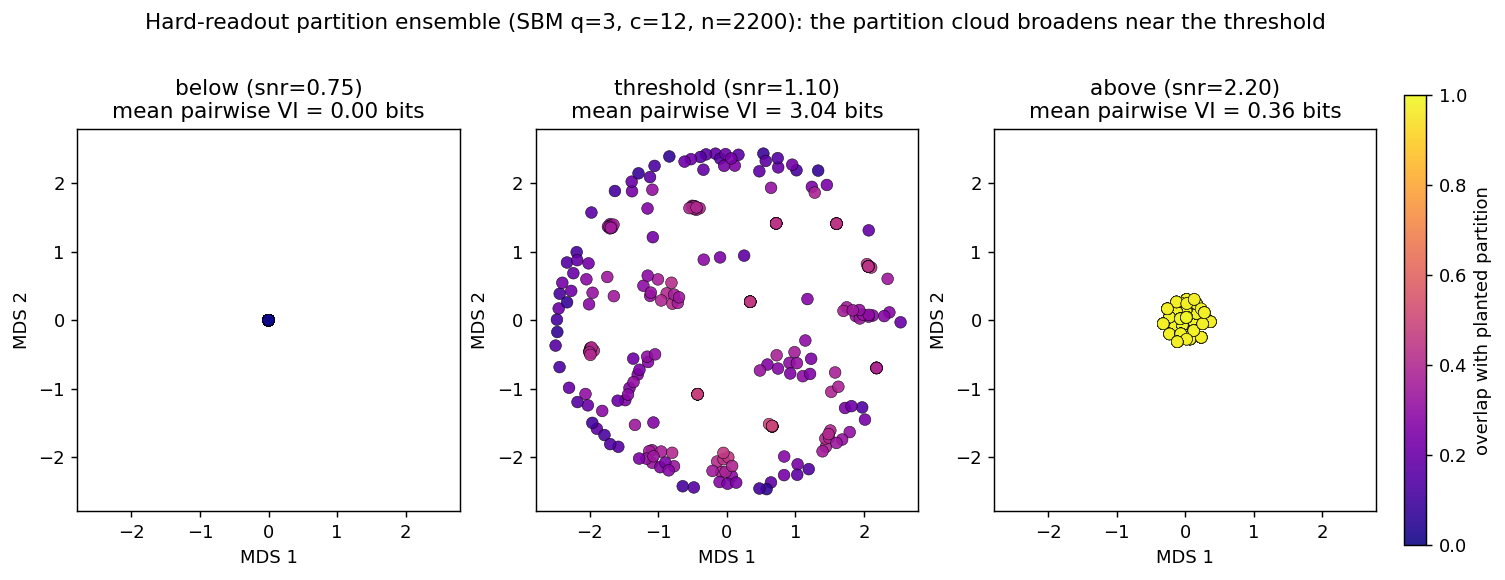}
\caption{The solution set broadens near the threshold. Each point is one detected
partition, placed by multidimensional scaling of pairwise variation of
information, colored by overlap with the planted partition. Below threshold a
single trivial solution; near the threshold a broad cloud of distant solutions;
deep above threshold a smaller consensus basin.}
\label{fig:basin}
\end{figure}

The mechanism behind the hard-readout peak is the multiplicity of detected partitions, which we
can see directly. We embed an ensemble of detected partitions in two dimensions by
multidimensional scaling of their pairwise variation of information
(Figure~\ref{fig:basin}). Deep below threshold the ensemble is a single point: the
only solution is the trivial one. Deep above threshold the ensemble contracts to a
small consensus basin. Near the threshold the ensemble broadens into many
mutually distant solutions, and the mean pairwise distance is largest there: in
the expanded landscape run the mean VI is $0.00$ below threshold, $3.05$ at
$\snr=1.10$, and $0.36$ at $\snr=2.20$. This geometric picture is consistent with
the broader statistical-physics view of metastable community-detection
landscapes, including the modularity complexity calculation of Kawamoto and
Kabashima~\cite{kawamoto2019}, while remaining separate from a direct computation of that
complexity.

\FloatBarrier

\section{Controls}
\label{sec:controls}

Two parts of the measurement procedure can create an apparent peak: pooling graph
samples with random initializations, and the hard cutoff that maps weakly
polarized BP marginals to the trivial partition. We first ran a smaller control
sweep at $n=2500$, $q=2$, and $c=10$ on the grid
$0.85\le\snr\le1.35$ with spacing $0.05$. The within-graph experiment computes
VI only among independent BP restarts on the same graph, then averages over
graphs. Its peak is at $\snr^\star=1.07$ with maximum VI $0.97$ bits. The pooled
graph-plus-initialization ensemble peaks at the same interpolated location,
$\snr^\star=1.07$, with maximum VI $1.79$ bits and a broader detectable-side
tail. Thus graph-to-graph sampling increases the amplitude and width of the
measured instability, while fixed-graph random restarts already produce a
near-threshold hard-readout peak.

The polarization cutoff affects the finite-size offset. To isolate this
dependence, we ran BP once at each $\snr$ with raw argmax output and then reread
the same outputs using cutoffs from $0.001$ to $0.1$. The peak locations are
$1.100$, $1.100$, $1.047$, $1.051$, $1.100$, $1.100$, and $1.128$ for cutoffs
$0.001,0.002,0.005,0.01,0.02,0.05,0.1$. The largest cutoff also lowers the peak
VI from about $1.83$ to $1.49$ bits. A log-cutoff fit of the peak location has
$R^2=0.12$, so the peak itself is a poor object for a cutoff law. The activation
point is much cleaner. Define $\snr_{50}(\tau)$ as the first $\snr$ where at
least half of the raw BP outputs have polarization above cutoff $\tau$. On the
same data,
\begin{equation}
\snr_{50}(\tau)-1 = 0.0086 + 0.522\,\tau,\qquad R^2=0.996 .
\end{equation}
We therefore correct for cutoff through the activation coordinate
$\snr-(\snr_{50}(\tau)-1)$ when comparing hard-readout curves across cutoffs.
This separates the engineered readout threshold from the BP fixed-point
dispersion measured by the residual and marginal-field controls.

We then measured the readout artifact directly on the same BP marginal fields.
For $\snr=0.85,0.90,0.95$, and $1.00$, the cutoff VI is exactly zero and the
cutoff-free marginal dispersion is at most $9.4\times10^{-6}$ bits per node. The
raw argmax hard readout from those same fields has VI $1.98$, $1.84$, $1.92$, and
$1.76$ bits, while independent samples from the marginals have VI
$1.997$--$1.998$ bits. The zero baseline below threshold is therefore a readout
convention, and the subthreshold BP field itself is uniform. As a structural
negative control, the same pipeline on a degree-heterogeneous core-periphery
family has maximum hard-readout VI $0.067$ bits, and the Erdos--Renyi endpoint has
zero VI.

We also instrumented the CPU BP reference to record iteration counts, the final
damped update size, an undamped fixed-point residual, the polarization, and a
Bethe free-entropy proxy for the BP factorization. A matched-seed budget control
at $100$, $300$, and $700$ iterations shows strong critical-slowing sensitivity:
at $\snr=1.05$, mean VI changes from $1.15$ bits to $1.70$ and $1.35$ bits, with
zero runs satisfying the strict $10^{-5}$ tolerance; at $\snr=1.10$, the values
are $1.77$, $1.64$, and $1.58$ bits, with one third of the $700$-iteration runs
passing the tolerance. The same-initialization VI between the $300$- and
$700$-iteration readouts is $0.98$ bits at $\snr=1.05$, $0.75$ bits at
$\snr=1.10$, and $0.32$ bits at $\snr=1.15$, then falls to zero by
$\snr=1.20$.

The long-budget residual gate in Eq.~\eqref{eq:residual-gated} gives the
cleanest separation. With matched initialization seeds at $700$ and $2500$
iterations, $\snr=1.00$ has zero VI and all $2500$-iteration runs pass the gate.
At $\snr=1.05$ and $1.10$, the $2500$-iteration hard VI is still $1.49$ and
$1.58$ bits, but only one third of runs pass the residual gate and the gated
subsets have zero VI and essentially zero marginal dispersion. At
$\snr=1.15,1.20,1.25$, and $1.30$, all or nearly all runs pass the gate; the
gated VI values are $1.31$, $1.30$, $1.29$, and $1.24$ bits, with
$D_{\Psi,\epsilon}^{\mathrm{fp}}$ equal to $0.056$, $0.069$, $0.075$, and
$0.085$ bits per node. The fixed-point ensemble in this experiment therefore
starts around $\snr=1.15$. The hard VI peak near $\snr=1.05$--$1.10$ combines
readout and relaxation effects; the budget-stable fixed-point dispersion lies
deeper in the detectable phase.
\FloatBarrier

\section{\texorpdfstring{A $q=5$ hard-phase probe}{A q=5 hard-phase probe}}
\label{sec:hard}

The continuous $q\le4$ regime covers only part of the statistical-physics
structure of the symmetric SBM. For $q\ge5$, the transition is first-order and an
informative fixed point can coexist with the uninformative one. A random-start BP
ensemble alone can therefore miss the hard phase. We added a targeted $q=5$
probe at $n=1400$, $c=20$, comparing the random BP initialization used elsewhere
with a weak planted warm start on the same graphs. The warm start serves as a
fixed-point probe.

This probe finds no well-converged sub-Kesten--Stigum coexistence band
for the implemented detector. For $\snr=0.70,0.775,0.85,0.925$, both random and
warm starts converge to the trivial output with zero overlap. At $\snr=1.00$,
the random-start overlap is $0.051$ and the warm-start overlap is $0.099$, but
the warm-start convergence fraction is only $0.33$. At $\snr=1.075$, the
overlaps are $0.201$ and $0.236$, with no runs passing the strict convergence
tolerance. At $\snr=1.15$, both initializations converge and reach overlap
$0.61$. Thus the present detector ensemble leaves the hard phase outside its
mapped regime and shows where this paper's continuous-transition analysis stops. Extending
partition-dispersion measurements into the hard phase would require a separate
experiment built around initialized BP, posterior sampling, or a 1RSB-style
state-counting calculation, beyond the random-start readout used for
$q\le4$.
\FloatBarrier

\section{Real-network label-free benchmarks}
\label{sec:real}

On a real network the block-model parameters are unknown, so we detect with a
degree-corrected Bethe-Hessian, the parameter-free spectral method that places its
informative eigenvalues below the bulk only when community structure is present and
reports an indivisible network otherwise~\cite{saade2014,newman2006eig}. We use
two real-data protocols. The first is the political-blogs network of Adamic and
Glance~\cite{adamic2005}, whose nodes carry a known political leaning; on its
giant component ($n=1222$, $16{,}714$ edges) the detector recovers the leaning
with overlap $0.86$. The second uses six public SNAP graphs spanning
organizational email, voting, collaboration, peer-to-peer, ego-social, and trust
networks: email-Eu-core ($n=986$, $m=16064$), Wiki-Vote ($n=7066$, $m=100736$),
CA-GrQc ($n=4158$, $m=13422$), p2p-Gnutella08 ($n=6299$, $m=20776$), Facebook
combined ($n=4039$, $m=88234$), and Epinions ($n=75877$, $m=405739$).

We then walk the network across its own spectral margin by degree-preserving
double-edge swaps, which weaken the community signal while holding the degree
sequence fixed. At each dilution level we compute two label-free quantities: the
edge-bootstrap hard-readout VI and the Bethe-Hessian modularity margin, defined as
the best modularity among sign splits induced by negative Bethe-Hessian
eigenvectors. Political labels are used only afterward to report recovery
overlap. We average over five independent rewiring realizations at each dilution
level. The hard-readout VI is near zero deep in the diluted regime, where the
detector reports an indivisible graph, rises to a peak when the unlabeled margin
is about $0.27$, and declines as the original structure becomes clearly
detectable.
The recovery of the political labels drops over the same range: at the
hard-readout peak the mean overlap is about $0.50$, whereas the undiluted graph
has overlap about $0.84$ under the same subsampling protocol. The label-derived
SNR is stored only as a calibration column in the ancillary results; the
horizontal coordinate of this experiment is the unlabeled spectral margin.

For the six SNAP graphs we avoid ground-truth labels entirely. We use nine
edge-keep probabilities from $1.00$ down to $0.05$, form an edge-subsampling
ensemble at each level, and compute Bethe-Hessian partitions. Each readout is
gated by a Chung--Lu null: the best Bethe-Hessian modularity must exceed the mean
plus two standard deviations of null graphs with the same expected degree
sequence. This adds an explicit real-network null distribution to the label-free
coordinate.

The six-network benchmark gives a clear boundary for the SBM interpretation. The
measurement is reproducible and label-free on all six graphs, with peak VI in the
range $1.94$--$2.00$ bits. Three graphs eventually lose null-significant structure
under heavy subsampling: at $p_{\rm keep}=0.05$, email-Eu-core has trivial
fraction $0.25$, CA-GrQc has $0.62$, and p2p-Gnutella08 has $0.75$. Wiki-Vote,
Facebook combined, and Epinions retain null-significant Bethe-Hessian structure
even at $p_{\rm keep}=0.05$. Thus real heterogeneous networks validate the
label-free computation and also show a limit of direct SBM transfer: the
real-network margin can remain high because degree heterogeneity, hubs, and
overlapping social structure generate spectral modes far outside the
Chung--Lu bulk.
\FloatBarrier

\section{Discussion}

\paragraph{Interpretation.} The measurement is a decomposition of detector-output
instability for a specified algorithmic pipeline. In the symmetric SBM the
hard-readout VI maximum lies in a weakly detectable band above the
Kesten--Stigum onset. The residual-gated control shifts the fixed-point claim
deeper into the detectable phase, beginning near $\snr=1.15$ in the long-budget
experiment. The Kesten--Stigum point remains the linear onset; the hard-readout
maximum is a finite-size detector curve. The measurement can complement
score-based model selection in this pipeline-specific sense: a high-scoring
partition can still be unstable under the specified restarts and perturbations. A
structural estimate of distance to threshold is still needed for a self-contained
diagnostic on unlabeled networks.

\paragraph{Scope.} The synthetic results are stated for the symmetric block model
with $q\le4$, where the transition is continuous and belief propagation is
threshold-optimal. For $q\ge5$ the model has a first-order hard phase between the
algorithmic and information-theoretic thresholds. The $q=5$ warm-start probe in
Section~\ref{sec:hard} found zero well-converged coexistence band for this
random-start readout, placing the hard phase outside the main claim. The
hard-readout VI is a property of the detector ensemble and the network together; a
different detector can trace a different instability band. The measurement also
differs from Bayesian posterior uncertainty: posterior samplers
define a principled distribution over partitions, whereas the BP ensemble used
here is an operational distribution induced by random initialization, graph
sampling, cutoff, and finite iteration budget. Comparing these two distributions
requires a separate posterior-sampling experiment.

The real-network experiments use label-free Bethe-Hessian coordinates in place of
SBM parameters. Political blogs supplies a labeled recovery check; the six SNAP
graphs supply a broader unlabeled benchmark with a Chung--Lu null. This benchmark
shows that the computation scales beyond a single graph, while also showing that
heterogeneous networks can preserve null-significant spectral modes deep into
edge subsampling. The real-network coordinate should therefore be read as a
spectral calibration coordinate, separate from an SBM parameter estimate. The
collapse metric now has a simple permutation null. The evidence supports
nonrandom finite-grid alignment; a limiting collapse function would require a
larger asymptotic study. The new large-$N$ audit reaches $N=100000$ with thirty
graph draws per sampled point. The measured high-$N$ scaling is a plateau
$\snr^\star-1\simeq0.024$, and the zero-asymptote exponent fit is unsupported by
the replicated data. The new convergence diagnostics show that the near-onset
hard VI peak mixes with critical slowing of BP under a finite iteration budget.
They also show a budget-stable fixed-point part of the curve:
from $\snr=1.15$ through $1.30$ in the 2500-step control, all or nearly all runs
pass the residual gate and the gated VI remains above $1.24$ bits. The Bethe
quantity recorded here is a free-entropy proxy for the implemented BP
factorization, below a calibrated posterior evidence. A full separation of
posterior metastability, graph-sampling variation, and algorithmic relaxation
dynamics requires additional experiments.

\section{Conclusion}

Hard-readout partition degeneracy, the disagreement among plausible partitions
returned by a specified community-detection pipeline, peaks inside the detectable
phase in the tested sparse symmetric SBMs. The peak is offset from the
Kesten--Stigum onset and appears as a broadening of the partition cloud. The
readout controls identify the measured object more sharply: uniform BP fields can
produce nearly $2$ bits of hard-label VI if labels are forced, while their
cutoff-free marginal dispersion is zero. The cutoff sweep shows protocol
dependence across cutoffs $0.001$--$0.1$, with peak locations from $1.047$ to
$1.128$. The long-budget residual gate separates critical slowing from
fixed-point dispersion: $\snr=1.05$ and $1.10$ have high hard VI but zero gated
VI, whereas $\snr=1.15$--$1.30$ retain gated VI $1.31$--$1.24$ bits. The
finite-size analysis now reaches $N=100000$ and reports measured offsets only;
the replicated large-$N$ offsets form a small plateau
$\snr^\star-1\simeq0.024$ with graph-bootstrap interval $[0.0227,0.0316]$. A
comparable instability appears in political blogs when the dilution is indexed by
an unlabeled Bethe-Hessian modularity margin. Six SNAP graphs extend the
real-network benchmark: the label-free computation is reproducible, while
heterogeneous networks can keep null-significant spectral structure even under
heavy edge subsampling. The $q=5$ probe is a negative result for this readout,
keeping the main claim in the continuous $q\le4$ regime. All reported quantities
are dimensionless and independent of the implementation backend.

\paragraph{Reproducibility.} All code and the raw numbers behind every figure are
in the ancillary files.

\end{document}